\documentclass[%
preprint,
 amsmath,amssymb,
 aps,
]{revtex4-1}

\usepackage{graphicx}
\usepackage{dcolumn}
\usepackage{amsmath}
\usepackage{amssymb}
\usepackage{bm}
\usepackage{slashed}
\usepackage{hyperref}
\usepackage{mathrsfs}

\usepackage{bm}

\def\p{\mbox{\boldmath$\displaystyle\mathbf{p}$}}

\def\0{\mbox{\boldmath$\displaystyle\mathbf{0}$}}
\def\s{\mbox{\boldmath$\displaystyle\mathbf{\sigma}$}}
\def\J{\mbox{\boldmath$\displaystyle\mathbf{J}$}}

\def\x{\mbox{\boldmath$\displaystyle\mathbf{x}$}}

\begin{document}


\title{Quantum field theory with a preferred direction: \\
The very special relativity framework}

\author{Cheng-Yang Lee}
\email{cylee@ime.unicamp.br}
\affiliation{
Institute of Mathematics, Statistics and Scientific Computation,\\
Unicamp, 13083-859 Campinas, S\~{a}o Paulo, Brazil
}%

\date{\today}

\begin{abstract}
The theory of very special relativity (VSR) proposed by Cohen and Glashow contains an intrinsic preferred direction. Starting from the irreducible unitary representation of the inhomogeneous VSR group $ISIM(2)$, we present a rigorous construction of quantum field theory with a preferred direction. We find, although the particles and their quantum fields between the VSR and Lorentz sectors are physically different, they share many similarities. The massive spin-half and spin-one vector fields are local and satisfy the Dirac and Proca equations respectively. This result can be generalised to higher-spin field theories. By studying the Yukawa and standard gauge interactions, we obtain a qualitative understanding on the effects of the preferred direction. Its effect is manifest for polarised processes but are otherwise absent.

\end{abstract}

\maketitle

\section{Introduction}
One of the important features of special relativity is that it provides an explanation to the null result of the Michelson-Morley experiment. Since its proposal by Einstein, special relativity and the underlying symmetry described by the Lorentz group has become one of the most important concepts in modern physics. 

The theory of VSR proposed by Cohen and Glashow provides a new perspective to the standard paradigm~\cite{Cohen:2006ky}. The theory, whose underlying symmetry group is a proper subgroup of the Lorentz group, is able to explain the null result of the Michelson-Morley experiment and reproduces many of the predictions of special relativity such as time dilation and length contraction. Therefore, there is a possibility that in the low energy limit, the symmetry of nature is described by VSR.

This hypothesis opens up a new direction to explore possible signatures of Lorentz violation. An advantage of this framework is that while Lorentz symmetry is violated, there remains a well-defined symmetry in which the theory must satisfy. There has been many works exploring extensions and modifications to quantum field theories and the Standard Model of particle physics (SM) by introducing Lorentz-violating but VSR-invariant interactions~\cite{Cohen:2006ir,Cohen:2006sc,Vohanka:2011aa,Maluf:2014yca,Alfaro:2013uga,Alfaro:2013uva,Upadhyay:2015yyt,
Bufalo:2015gja,Alfaro:2015fha,Nayak:2015xba}. In some of these works, it has been proposed that VSR-invariant fields have different kinematic to their Lorentz-invariant counterparts. The main motivation for most of these works is due to the fact that the VSR symmetry contains an intrinsic preferred direction which is characterised by the existence of an invariant null vector. By studying interactions involving the null vector, one can obtain information on the effects of the preferred direction and signatures of Lorentz-violation.

In this paper, we take a different approach. An important point to note is that the kinematics of Lorentz-invariant theory does not require any modifications. One is free to introduce additional VSR-invariant interactions but we will not consider this possibility here. 

Starting from the irreducible unitary representation of $ISIM(2)$, we provide a rigorous construction of VSR-invariant quantum fields. The solutions to the free fields are completely determined, up to a global constant. For the fermionic and vector field, it is shown that they are local and satisfy the Dirac and Proca equation respectively. This result can be generalised to higher-spin where the locality and field equations between the VSR and Lorentz sector are identical.

We propose the possibility that particles and their quantum fields with VSR symmetry are physically distinct from the SM particles. Therefore, their existence would constitute new physics beyond the SM. This proposal seems quite  obvious as particles with VSR symmetry must furnish the irreducible unitary representation of $ISIM(2)$ and not $ISO(3,1)$. 

Although quantum fields with VSR and Lorentz symmetry have many similarities, as we will show in sec.~\ref{sec:qf}, there remain important qualitative differences. Restricting ourselves to the Yukawa and standard gauge interactions, we show that for unpolarised processes, the resulting observables are identical to their Lorentz-invariant counterparts. The processes in which the results between the two sectors differ and where the effect of the preferred direction is manifest, are the polarised processes.

The paper is organised as follows. Section~\ref{Sec:VSR-Review} offers a brief review on the VSR groups and their transformations. In sec.~\ref{Casimir_invariants_VSR}, we derive the particle states and their transformations from the irreducible unitary representations of $ISIM(2)$. The subsequent section provides a rigorous construction of VSR-invariant quantum field theory. The massive fermionic and vector fields are constructed as examples.
\section{The VSR group and their transformations}\label{Sec:VSR-Review}
In this section, we provide a brief review of the VSR transformations.
The VSR groups are generated from the algebras summarised in tab.~\ref{table:1} where
\begin{equation}
T^{1}=K^{1}+J^{2},\hspace{0.5cm} T^{2}=K^{2}-J^{1}.\label{eq:T1T2}
\end{equation}
They are proper Lorentz subgroups and their transformations satisfy the postulates of special relativity, namely the existence of a maximal velocity invariant in all inertial frames~\cite{Horvath}.

%

\begin{table}[!hbt]
\centering
\begin{tabular}{lll}
\toprule
Algebra & \hspace{1cm} Generators\\
\hline
$\mathfrak{t}(2)$    & \hspace{1cm} $T^1,T^2$ \\
$\mathfrak{e}(2)$    & \hspace{1cm} $T^1,T^2,J^3$ \\
$\mathfrak{hom}(2)$  & \hspace{1cm} $T^1,T^2,K^3$ \\
$\mathfrak{sim}(2)$  & \hspace{1cm} $T^1,T^2,J^3,K^3$ \\
\toprule
\end{tabular}
\caption{\label{table:1}The four VSR algebras.}
\end{table}

There are two properties of VSR that are important. Firstly, the inclusion of  any of the following 
discrete symmetries $\mathsf{P}$, $\mathsf{T}$, $\mathsf{CP}$ or $\mathsf{CT}$ 
will yield the full Lorentz group. This suggests that the VSR-invariant quantum fields may violate
the mentioned discrete symmetries. But as long as the fields are local, $\mathsf{CPT}$ should be conserved. Secondly, all the VSR algebras have a preferred direction. In this paper, we chose the preferred direction to coincide with the 3-axis. After reviewing the $SIM(2)$ transformations, we will derive the transformations of the particle states and quantum fields with VSR symmetry.

The VSR transformations are generated by $\mathfrak{sim}(2)$ where the finite-dimensional generators are obtained using eq.~(\ref{eq:T1T2}) in the vector representation. They consist of rotations and boosts. The former is identical to the rotation in the Lorentz group along the 3-axis and the later is defined as a product of all three group elements
\begin{eqnarray}
L(p)&=&T_{1}(\beta_{1})T_{2}(\beta_{2})L_{3}(\varsigma) \nonumber\\
&=&e^{i\beta_{1}\mathscr{T}^{1}}e^{i\beta_{2}\mathscr{T}^{2}}e^{i\mathscr{K}^{3}\varsigma}\label{eq:VSR_boost}
\end{eqnarray}
where $\mathscr{T}^{i}$, $\mathscr{K}^{3}$ are the VSR generators in the vector representations and the parameters are given by
\begin{eqnarray}
&&\beta_{1}=\frac{p^{1}}{p^{0}-p^{3}},\\
&&\beta_{2}=\frac{p^{2}}{p^{0}-p^{3}},\\
&&\varsigma=-\ln\left(\frac{p^{0}-p^{3}}{m}\right).\label{eq:vsr_parameters}
\end{eqnarray}
The boost is defined such that it takes $k^{\mu}=(m,\0)$ to arbitrary momentum $p^{\mu}=(p^{0},\p)$ where $p^{0}=\sqrt{|\p|^{2}+m^{2}}$.

\section{Particle states}\label{Casimir_invariants_VSR}
An important feature of the Poincar\'{e} algebra is that the eigenvalues of the Casimir operators have the physical interpretation of mass and spin. Similarly, the particle states of VSR are also defined as simultaneous eigenstates of the Casimir invariants of $\mathfrak{isim}(2)$~\cite[tab.VII, row 7]{Patera:1975bw}
\begin{equation}
C_{1}=P^{\mu}P_{\mu},\hspace{0.5cm}
C_{2}=J^{3}-\frac{P^{2}}{P^{0}-P^{3}}T^{1}+\frac{P^{1}}{P^{0}-P^{3}}T^{2}. \label{eq:Casimir_VSR}
\end{equation}
Mass remains a valid description of VSR particles but the notion of spin is different. To see this, consider the one-particle state $|k,\sigma\rangle$ where $k^{\mu}$ is the momentum vector given in tab.~\ref{table:2} and $\sigma$ represents all other continuous or discrete degrees of freedom. Since $[P^{2},T^{1}]=[P^{1},T^{2}]=0$, the second and third terms of $C_{2}$ identically vanish upon acting on $|k,\sigma\rangle$. Therefore, the spectrum of $C_{2}$ are the eigenvalues of $J^{3}$~\footnote{This was brought to my attention by D.~V.~Ahluwalia.}.

Let $\Lambda$ be an element of $SIM(2)$ and $U(\Lambda)$ be its linear and unitary representation. Since $\mathfrak{isim}(2)$ is a sub-algebra of the the Poincar\'{e} algebra, $P^{\mu}$ still transforms as
\begin{equation}
U(\Lambda)P^{\mu}U^{-1}(\Lambda)=\Lambda_{\nu}^{\,\,\mu}P^{\nu}.
\end{equation}
Therefore, the one-particle has the following transformation
\begin{equation}
U(\Lambda)|p,\sigma\rangle=\sum_{\sigma'}C_{\sigma'\sigma}(\Lambda,p)|\Lambda p,\sigma'\rangle
\label{eq:general_transs}.
\end{equation}
Since $U(\Lambda)$ is unitary, depending on the definition of the inner-product of the particle state, $C_{\sigma'\sigma}(\Lambda,p)$ must furnish finite-dimensional unitary representation of $SIM(2)$ up to a normalisation factor. In the following sections, we will determine $C_{\sigma'\sigma}(\Lambda,p)$.

\begin{table}[!hbt]
\centering
\begin{tabular}{lll}
\toprule
 & Standard $k^{\mu}$ & Little group \\
\hline
$p^{\mu}p_{\mu}=m^{2}$, $p^{0}>0$ & $(m,0,0,0)$ & $SO(2)$ \\
$p^{\mu}p_{\mu}=0$, $p^{0}>0$ & $(\kappa,0,0,\kappa)$ & $E(2)$\\

\toprule
\end{tabular}
\caption{The little groups}
\label{table:2}
\end{table}

\subsection{Massive particle state}
The rotation and boost generators of $SIM(2)$ along the 3-axis are identical to their Lorentz counterparts while $T^{1}$ and $T^{2}$ can be obtained from eq.~(\ref{eq:T1T2}). From these generators, the only non-trivial unitary representation of $SIM(2)$ is the rotation about the 3-axis. Therefore, the solutions to $C_{\sigma'\sigma}(\Lambda,p)$ are given by
\begin{eqnarray}
&& C_{\sigma'\sigma}(L(p))=N(p)\delta_{\sigma'\sigma},\label{eq:trivial} \\
&& C_{\sigma'\sigma}(R_{3}(\phi))=\exp(i J_{\sigma'\sigma}^{3}\phi)
\end{eqnarray}
where $N(p)$ is a normalisation factor to be determined. The unitary finite-dimensional representation of $J^{3}$ is chosen to be $(J^{3})_{\sigma'\sigma}=\sigma\delta_{\sigma'\sigma}$ with $\sigma=-j,\cdots,j$ where $j$ takes either integer or half-integer. For a general one-particle state $|p,\sigma\rangle$, we define the following orthogonal inner-product
\begin{equation}
\langle p',\sigma'|p,\sigma\rangle=\delta_{\sigma'\sigma}\delta^{3}(\p'-\p).\label{eq:ip}
\end{equation} 
The normalisation factor is then fixed to $N(p)=(p^{0}/m)^{1/2}$~\footnote{A detailed derivation for the normalisation factor can be found in~\cite[sec.2.5]{Weinberg:1995mt}}. Therefore, the effect of a boost on the massive particle state is
\begin{equation}
|p,\sigma\rangle=\sqrt{\frac{m}{p^{0}}}U(L(p))|k,\sigma\rangle.\label{eq:boost}
\end{equation}
From eq.~(\ref{eq:boost}), we may then use the method of induced representation~\cite{Weinberg:1995mt} to obtain the VSR transformation of the one-particle state
\begin{equation}
U(\Lambda)|p,\sigma\rangle=\sqrt{\frac{(\Lambda p)^{0}}{p^{0}}}\sum_{\sigma'}
D_{\sigma'\sigma}(W(\Lambda,p))|\Lambda p,\sigma'\rangle.\label{eq:trans_particle}
\end{equation}
This equation takes the same form as the Lorentz transformation of a massive one-particle state, except the little group element $W(\Lambda,p)$ is defined as
\begin{equation}
W(\Lambda,p)=L^{-1}(\Lambda p)\Lambda L(p)
\end{equation}
where $\Lambda\in SIM(2)$ is an arbitrary VSR transformation. In VSR, the elements of the little group for massive particle are elements of $SO(2)$. In this paper, this corresponds to rotations about the 3-axis. 

The group is infinitely connected $SO(2)\sim R/Z_{\infty}$ and its universal covering group is $R$, the group of real numbers under addition~\cite[sec.~16.24]{Wybourne}. Its multi-valued unitary irreducible representations are~\cite{Binegar:1981gv}
\begin{equation}
D_{\sigma'\sigma}(R(\phi))
=e^{i\sigma\phi}\delta_{\sigma'\sigma}
\end{equation}
where $\sigma$ is a real number that labels the representation. The irreducible representations are one-dimensional since $SO(2)$ is an Abelian group. Therefore, the transformation of the massive one-particle state is
\begin{equation}
U(\Lambda)|p,\sigma\rangle=\sqrt{\frac{(\Lambda p)^{0}}{p^{0}}}e^{i\sigma\phi(\Lambda,p)}|\Lambda p,\sigma\rangle.\label{eq:vsr_massive_trans_particle}
\end{equation}

\subsection{Massless particle state}

The little group for massless VSR particles is the Euclidean group $E(2)$ which is also the little group of massless particles in the Lorentz group. Here, a problem of concern is the kinematics of the massless particles. In the vector representation, the transformation $\exp(i\beta_{i}\mathscr{T}^{i})$, 
leaves $k^{\mu}=(\kappa,0,0,\kappa)$ invariant so the boost is
\begin{eqnarray}
L^{\mu}_{\,\,\,\nu}(p)k^{\nu}&=&(e^{i\mathscr{K}^{3}\varsigma})^{\mu}_{\,\,\,\nu}k^{\nu} \nonumber \\
&=&(p,0,0,p)
\end{eqnarray}
where the second line is obtained using the parameter $\ln\varsigma=p/\kappa$. This is obviously undesirable as the motion of massless particles are restricted to the spatial dimension coinciding with the preferred direction. From this simple observation, we conclude that VSR does not admit a physical description of massless particle. 

In the literature, there has been works studying the VSR-invariant electrodynamics where the starting point is a Lorentz-violating but manifestly VSR-invariant Lagrangian. Working under the hypothesis that the space-time symmetry is VSR, we are certainly allowed to introduce VSR-invariant interactions as long as they are gauge-invariant. However, one should not modify the underlying Lorentz-invariant kinematics since this implicitly implies that the photons furnish the irreducible unitary representation of $ISIM(2)$. But since VSR does not admit a physical description of massless particles, such constructs do not seem to be plausible.

\section{Massive quantum fields}\label{sec:qf}

Because of the problem with massless particles in VSR, we will only focus on the construction of massive quantum fields with VSR symmetry. Let $\psi^{(\sigma)}(x)$ be a quantum field describing a massive particle state $|p,\sigma\rangle$
\begin{equation}
\psi^{(\sigma)}(x)=(2\pi)^{-3/2}\int \frac{d^{3}p}{\sqrt{2p^{0}}}\Big[e^{-ip\cdot x}u(\p,\sigma)a(\p,\sigma)
+e^{ip\cdot x}v(\p,\sigma)b^{\dag}(\p,\sigma)\Big].\label{eq:qf}
\end{equation}
The field defined in eq.~(\ref{eq:qf}) is manifestly covariant under space-time translations. The demand of VSR-covariance means that it must transform as
\begin{equation}
U(\Lambda)\psi^{(\sigma)}_{\ell}(x)U^{-1}(\Lambda)= \mathcal{D}_{\ell\bar{\ell}}
(\Lambda^{-1})\psi^{(\sigma)}_{\bar{\ell}}(\Lambda x)\label{eq:vsr_qf_transform}
\end{equation}
where $\mathcal{D}(\Lambda)$ is the finite-dimensional representation of $SIM(2)$. Equation (\ref{eq:vsr_qf_transform}) and the transformations of the massive particle states given by
eq.~(\ref{eq:vsr_massive_trans_particle}) are sufficient for us to determine the coefficients and how they
transform. Following the same derivation given in~\cite[sec.~5.1]{Weinberg:1995mt}
the coefficients of arbitrary momentum are given by
\begin{equation}
u(\p,\sigma)=\mathcal{D}(L(p))u(\0,\sigma),\label{eq:vsr_u_boost}
\end{equation}
\begin{equation}
v(\p,\sigma)=\mathcal{D}(L(p))v(\0,\sigma)\label{eq:vsr_v_boost}
\end{equation}
and the coefficients at rest are determined by the demand of rotation symmetry
\begin{eqnarray}
&&\sum_{\bar{\sigma}}D_{\sigma\bar{\sigma}}(R)u_{\ell}(\0,\bar{\sigma})=
\sum_{\bar{\ell}}\mathcal{D}_{\ell\bar{\ell}}(R)
u_{\bar{\ell}}(\0,\sigma),\\
&&\sum_{\bar{\sigma}}D^{*}_{\sigma\bar{\sigma}}(R)v_{\ell}(\0,\bar{\sigma})=
\sum_{\bar{\ell}}\mathcal{D}_{\ell\bar{\ell}}(R)
v_{\bar{\ell}}(\0,\sigma).
\end{eqnarray}
Expanding the matrices about the identity with $\mathcal{D}(R)=1+i\mathcal{J}^{3}\phi$ and $D_{\sigma\bar{\sigma}}(R)=\delta_{\sigma\bar{\sigma}}(1+i\sigma\phi)$,
we find that the coefficients at rest are eigenvectors of $\mathcal{J}^{3}$
\begin{eqnarray}
&&\mathcal{J}^{3}\,u(\0,\sigma)=\sigma u(\0,\sigma),\\ \label{eq:eigenvectors1}
&&\mathcal{J}^{3}\,v(\0,\sigma)=-\sigma v(\0,\sigma). \label{eq:eigenvectors2}
\end{eqnarray}

The demand of locality requires the quantum fields to commute or anti-commute with its adjoint at space-like separation. The field $\psi^{(\sigma)}(x)$ given by eq.~(\ref{eq:qf}) does not satisfy this criteria. This problem can be resolved starting with the observation that the rotation generator $\mathcal{J}^{3}$ has a spectrum of $-j,\cdots,j$. Therefore, there exists $2j+1$ fields $\psi^{(-j)}(x),\cdots,\psi^{(j)}(x)$ that transform according to the same finite-dimensional representation even though each fields formally correspond to a different particle specie. Therefore, given a finite-dimensional representation $\mathcal{D}(\Lambda)$ of dimension $2j+1$,  we can construct a new VSR-invariant field $\psi(x)$ as the sum of all fields from $\psi^{(-j)}(x)$ to $\psi^{(j)}(x)$
\begin{equation}
\psi(x)=\sum_{\sigma=-j}^{j}\psi^{(\sigma)}(x).
\label{eq:vsr_psi}
\end{equation}
One could consider a more general linear combination $\sum_{\sigma}f_{\sigma}\psi^{(\sigma)}(x)$. But at the end, $f_{\sigma}$  must be fixed by the demand of locality and a positive-definite Hamiltonian.
In the subsequent sections, we construct the spin-half and spin-one vector fields and show that they are both local (provided that $f_{\sigma}=1$) and satisfy the Dirac and Proca equation respectively.

By construction, the field $\psi(x)$ is similar to a quantum field of spin-$j$ representation of the Lorentz group with the same rotation generator along the $3$-axis.  They have the same number of degrees of freedom and their coefficients at rest take the same form. In the subsequent sections, this correspondence allow us to compare the VSR and Lorentz-invariant theories.

\subsection{Spin-half representation}\label{vsr_spin_field}

In the Lorentz group, given the rotation generators $\J$, we can always find two solutions to the boost generators given by $\mathbf{K}_{\pm}=\pm i\J$ such that the Lorentz algebra is satisfied. The spin-half generators of the Lorentz group are $\J=\frac{1}{2}\s$, $\mathbf{K}_{\pm}=\pm\frac{1}{2}i\s$ where $\s=(\sigma^{1},\sigma^{2},\sigma^{3})$ are the Pauli matrices. Therefore, we can construct two sets of generators that satisfy the $\mathfrak{sim}(2)$ algebra
\begin{equation}
K^{3}_{\pm}=\pm iJ^{3},\hspace{0.5cm}
T^{1}_{\pm}=K^{1}_{\pm}+J^{2},\hspace{0.5cm}
T^{2}_{\pm}=K^{2}_{\pm}-J^{1}. \label{eq:vsr_spin_half_rep}
\end{equation}
which then allows us to construct the following four-dimensional representation~\cite{Ahluwalia:2010zn}
\begin{eqnarray}
&&\mathcal{J}^{3}=
\left(\begin{array}{cc}
J^{3} & O \\
O & J^{3}
\end{array}\right),\hspace{0.5cm}
\mathcal{K}^{3}=
\left(\begin{array}{cc}
K^{3}_{-} & O \\
O & K^{3}_{+}
\end{array}\right)\\
&&\mathcal{T}^{1}=
\left(\begin{array}{cc}
T^{1}_{-} & O \\
O & T^{1}_{+}
\end{array}\right),\hspace{0.5cm}
\mathcal{T}^{2}=
\left(\begin{array}{cc}
T^{2}_{-} & O \\
O & T^{2}_{+}
\end{array}\right). 
\end{eqnarray}

The field that transforms according to the above representation is given by eq.~(\ref{eq:qf}) with $\sigma=\pm\frac{1}{2}$.
By the demand of VSR-covariance, the coefficients at rest must be eigenvectors of $\mathcal{J}^{3}$ whose solutions can be taken to be
\begin{equation}
u(\0,\textstyle{\frac{1}{2}})=\sqrt{m_{\psi}}\left(\begin{array}{cccc}
1 \\
0\\
1 \\
0 \end{array}\right),\hspace{0.5cm}
u(\0,-\textstyle{\frac{1}{2}})=\sqrt{m_{\psi}}\left(\begin{array}{cccc}
0 \\
1 \\
0\\
1 \end{array}\right),\label{eq:c_coefficients}
\end{equation}
\begin{equation}
v(\0,\textstyle{\frac{1}{2}})=\sqrt{m_{\psi}}\left(\begin{array}{cccc}
0 \\
1\\
0\\
-1 \end{array}\right),\hspace{0.5cm}
v(\0,-\textstyle{\frac{1}{2}})=\sqrt{m_{\psi}}\left(\begin{array}{cccc}
-1\\
0\\
1\\
0 \end{array}\right)\label{eq:d_coefficients}.
\end{equation}
These solutions, up to a normalisation factor are in agreement with the Dirac spinors at rest for a Lorentz-invariant spin-half field. But for arbitrary momentum, the coefficients $u(\p,\sigma)$ and $v(\p,\sigma)$ are different from the Dirac spinors. This is best characterised by the following VSR-invariant equation in which they satisfy~\cite{Cohen:2006ir}
\begin{eqnarray}
&&\left(\gamma^{\mu}p_{\mu}-\frac{m_{\psi}^{2}}{2}\frac{\slashed{n}}{p\cdot n}\right)P_{\mp}u(\p,\pm\textstyle{\frac{1}{2}})=0,\\
&&\left(\gamma^{\mu}p_{\mu}-\frac{m_{\psi}^{2}}{2}\frac{\slashed{n}}{p\cdot n}\right)P_{\pm}v(\p,\pm\textstyle{\frac{1}{2}})=0
\end{eqnarray}
where $n^{\mu}=(1,0,0,1)$ is the VSR-invariant null vector and $P_{\pm}=(I\pm\gamma^{5})$ is the projection operator.
In the representation we are working with, the $\gamma^{\mu}$ is given by
\begin{equation}
\gamma^{0}=\left(
\begin{matrix}
O & I \\
I & O
\end{matrix}\right),\hspace{0.3cm}
\gamma^{i}=\left(
\begin{matrix}
O & -\sigma^{i} \\
\sigma^{i} & O
\end{matrix}\right),\hspace{0.3cm}
\gamma^{5}=\left(
\begin{matrix}
I & 0 \\
0 &-I
\end{matrix}\right).
\end{equation}
Since these equations are not polynomials in momentum, they cannot provide a local kinematics to the theory.
Additionally, because both $u(\p,\sigma)$ and $v(\p,\sigma)$ have the same mass term, it is impossible to write down a field equation in the configuration space. Fortunately, this turns out to be inconsequential. Explicit computation shows that $u(\p,\sigma)$ and $v(\p,\sigma)$ satisfy the Dirac equation in the momentum space
\begin{eqnarray}
&&(\gamma^{\mu}p_{\mu}-m_{\psi}I)u(\p,\sigma)=0,\\
&&(\gamma^{\mu}p_{\mu}+m_{\psi}I)v(\p,\sigma)=0.
\end{eqnarray}
They also have the spin-sums
\begin{eqnarray}
&&\sum_{\sigma}u(\p,\sigma)\overline{u}(\p,\sigma)(p)=(\gamma^{\mu}p_{\mu}-m_{\psi}I),\\
&&\sum_{\sigma}v(\p,\sigma)\overline{v}(\p,\sigma)(p)=(\gamma^{\mu}p_{\mu}+m_{\psi}I)
\end{eqnarray}
which are identical to their Lorentz-covariant counterparts where $\overline{u}(\p,\sigma)=u^{\dag}(\p,\sigma)\gamma^{0}$ and likewise for $\overline{v}(\p,\sigma)$. Therefore, from the field equations and spin-sums, we obtain a local VSR-invariant field of spin-half
\begin{equation}
\psi(x)=\sum_{\sigma=\pm1/2}\psi^{(\sigma)}(x)
\end{equation}
which satisfies the Dirac equation. The Lagrangian for $\psi(x)$ is simply
\begin{equation}
\mathscr{L}_{\psi}=\overline{\psi}(i\gamma^{\mu}\partial_{\mu}-m_{\psi}I)\psi.
\end{equation}
It is instructive to note that one could try to construct a theory for $\psi^{(\sigma)}(x)$ using the Dirac Lagrangian. However, this theory would not be physical. The main reason being that the field $\psi^{(\sigma)}(x)$ is non-local. Therefore, the resulting $S$-matrix is not VSR-invariant.

Although the VSR spin-half fields are physically different from the Dirac field, they have the same field equations and locality structure . Repeating the same construction for higher spin fields, we have shown that  up to spin-two, the field equations and locality structure of VSR and Lorentz-invariant fields are identical. This suggests these relations hold for arbitrary spin representation. Consequently, the coefficients at rest of a massive spin-$j$ quantum field with VSR symmetry can always be chosen such that the field is local and satisfies the same field equation as their Lorentz counterpart of the $(j,0)\oplus(0,j)$ representation.

In hindsight, this result is expected. By examining the proof for the existence of $t^{\mu_{1}\mu_{2}\cdots\mu_{2j}}$~\cite{Weinberg:1964cn}, we see that since $SIM(2)$ transformations are also Lorentz transformations, it follows that there must exists a symmetric traceless rank $2j$ tensor in which $\mathcal{D}(L(p))\mathcal{D}^{\dag}(L(p))$ may be expressed as for arbitrary spin. However, we did not prove the VSR tensor must be identical to $t^{\mu_{1}\mu_{2}\cdots\mu_{2j}}$.

\subsection{Vector representation}
Now we consider the real massive vector field. Here the coefficients at rest are eigenvectors of the rotation generator along the 3-axis in the vector representation so there are four solutions and they can be chosen to be
\begin{eqnarray}
&&\phi^{\mu}(\0,0)=\left(\begin{matrix}
-i \\
 0 \\
 0 \\
 0
\end{matrix}\right),
e^{\mu}(\0,1)=\frac{1}{\sqrt{2}}
\left(\begin{array}{cccc}
 0 \\
-1 \\
-i \\
 0 \end{array}\right),\\
&& e^{\mu}(\0,0)=
\left(\begin{array}{cccc}
0 \\
0 \\
0 \\
1 \end{array}\right),
e^{\mu}(\0,-1)=\frac{1}{\sqrt{2}}
\left(\begin{array}{cccc}
 0 \\
 1 \\
-i \\
 0 \end{array}\right).
\end{eqnarray}
The first solution at finite momentum is equal to $-ip^{\mu}$ so that the associated field is simply the derivative of a real scalar field $\partial^{\mu}\phi(x)$. The remaining solutions are identical to the polarisation vectors at rest for a Lorentz-invariant vector field. We associate them with
\begin{equation}
A_{\sigma}^{\mu}(x)=(2\pi)^{-3/2}\int\frac{d^{3}p}{\sqrt{2p^{0}}}\Big[e^{-ip\cdot x}e^{\mu}(\p,\sigma)c(\p,\sigma)+e^{ip\cdot x}e^{\mu*}(\p,\sigma)c^{\dag}(\p,\sigma)\Big].
\end{equation}
At finite-momentum, these vectors satisfy the familiar relation $p_{\mu}e^{\mu}(\p,\sigma)=0$ which translates to
$\partial_{\mu}A^{\mu}_{\sigma}(x)=0$ in the configuration space. The contraction $n_{\mu}e^{\mu}(\p,\sigma)$ only vanishes for $\sigma=\pm1$. The vectors $e^{\mu}(\p,\pm1)$ have well-defined massless limit but the field constructed from a sum of $A^{\mu}_{\pm1}(x)$ is non-local. A local vector is constructed by summing over all fields
\begin{equation}
A^{\mu}(x)=\sum_{\sigma=\pm1,0}A^{\mu}_{\sigma}(x).
\end{equation}
One can verify that the field commutes with its adjoint at space-like separation using the following spin-sum
\begin{equation}
\sum_{\sigma}e^{\mu}(\p,\sigma)e^{\nu*}(\p,\sigma)=-\eta^{\mu\nu}+\frac{p^{\mu}p^{\nu}}{m_{A}^{2}}.
\end{equation}
It should not come as a surprise that this is identical to its Lorentz-invariant counterpart. Given these results, the Lagrangian for the VSR-invariant vector field is
\begin{equation}
\mathscr{L}_{A}=-\frac{1}{4}F^{\mu\nu}F_{\mu\nu}+\frac{1}{2}m_{A}^{2}A^{\mu}A_{\mu}
\end{equation}
where $F_{\mu\nu}=\partial_{\mu}A_{\nu}-\partial_{\nu}A_{\mu}$ is the field strength tensor.

\subsection{Discrete symmetries}

An important difference between VSR and Lorentz-invariant quantum fields is the discrete symmetries. The VSR algebras cannot accommodate discrete symmetry operators $\mathsf{P}$, $\mathsf{T}$, $\mathsf{CP}$ and $\mathsf{CT}$. Including any one of them would yield the full Lorentz algebra. Therefore, it is expected that the VSR-invariant fields would violate these symmetries but preserve charge-conjugation.

Here we consider the discrete symmetries of the massive spin-half and vector fields constructed in the 
previous section. Although VSR algebra does not admit $\mathsf{P}$ and $\mathsf{T}$
generators, it does not prevent us from studying their discrete transformations.The actions of $\mathsf{P}$ and $\mathsf{T}$ on the creation and annihilation operators are given by
\begin{eqnarray}
&&\mathsf{P}a(\p,\sigma)\mathsf{P}^{-1}=\eta^{*} a(-\p,\sigma),\\
&&\mathsf{P}b(\p,\sigma)\mathsf{P}^{-1}=\bar{\eta}^{*} b(-\p,\sigma),\\
&&\mathsf{T}a(\p,\sigma)\mathsf{T}^{-1}=\varrho^{*}(-1)^{j-\sigma}a(-\p,-\sigma),\\
&&\mathsf{T}b(\p,\sigma)\mathsf{T}^{-1}=\bar{\varrho}^{*}(-1)^{j-\sigma}b(-\p,-\sigma)
\end{eqnarray}
where $\eta$ and $\varrho$ are the parity and time-reversal phases for particles and
$\bar{\eta}$ and $\bar{\varrho}$ are the parity and time-reversal phases for
anti-particles. Under charge-conjugation
\begin{eqnarray}
&&\mathsf{C}a(\p,\sigma)\mathsf{C}^{-1}=\varsigma^{*}b(\p,\sigma),\\
&&\mathsf{C}b(\p,\sigma)\mathsf{C}^{-1}=\bar{\varsigma}^{*}a(\p,\sigma)
\end{eqnarray}
where $\varsigma$ and $\bar{\varsigma}$ are the charge-conjugation phases.

Table~\ref{table:3} provides the discrete symmetry transformations for $\psi(x)$ and $A^{\mu}(x)$. It shows that charge-conjugation is a symmetry. When the fields are restricted to the preferred direction, parity and time-reversal which are otherwise violated, are conserved.  This is in agreement with the original
observation of Cohen and Glashow that VSR violates $\mathsf{P}$, $\mathsf{T}$, $\mathsf{CP}$. If we consider the action of $\mathsf{CPT}$, we find that it is a symmetry of the theory
\begin{eqnarray}
&&(\mathsf{CPT})\psi(x)(\mathsf{CPT})^{-1}=-(\eta_{\psi}\varrho_{\psi}\varsigma_{\psi})^{*}\gamma^{5}
\psi^{*}(-x),\\
&&(\mathsf{CPT})A^{\mu}(x)(\mathsf{CPT})^{-1}=-(\eta_{A}\varrho_{A}\varsigma_{A})A^{\mu}(-x).
\end{eqnarray}

\begin{table*}
\begin{ruledtabular}
\begin{tabular}{llll}
\multicolumn{1}{l}{Fermionic field}&
&
\multicolumn{1}{l}{Vector field}&
\\
\colrule
$\mathsf{C}\psi(t,\x)\mathsf{C}^{-1}=i\varsigma_{\psi}\gamma^{2}\psi^{*}(t,\x)$ & $\varsigma_{\psi}^{*}=\bar{\varsigma}_{\psi}$ & 
$\mathsf{C}A^{\mu}(t,\x)\mathsf{C}^{-1}=\varsigma_{A}A^{\mu}(t,\x)$ & $\varsigma_{A}^{*}=\varsigma_{A}$\\
$\mathsf{P}\psi(t,x^{3})\mathsf{P}^{-1}=\eta_{\psi}^{*}\gamma^{0}\psi(t,-x^{3})$ &
$\eta^{*}_{\psi}=-\bar{\eta}_{\psi}$ &
$\mathsf{P}A^{\mu}(t,x^{3})\mathsf{P}^{-1}=-\eta_{A}P^{\mu}_{\,\,\nu}A^{\nu}(t,-x^{3})$ &
$\eta_{A}=\eta^{*}_{A}$\\
$\mathsf{T}\psi(t,x^{3})\mathsf{T}^{-1}=\varrho^{*}i\gamma^{0}\gamma^{2}\gamma^{5}\psi(-t,x^{3})$&
$\varrho^{*}_{\psi}=\bar{\varrho}_{\psi}$ &
$\mathsf{T}A^{\mu}(t,x^{3})\mathsf{T}^{-1}=-\varrho_{A}T^{\mu}_{\,\,\nu}A^{\nu}(-t,x^{3})$ &
$\varrho_{A}=\varrho_{A}^{*}$
\end{tabular}
\end{ruledtabular}
\caption{\label{table:3}
Discrete symmetries for $\psi(x)$ and $A^{\mu}(x)$ where 
$P^{\mu}_{\,\,\nu}=-T^{\mu}_{\,\,\nu}=\mbox{diag}(1,-1,-1,-1)$. Parity and time-reversal are symmetries of the fields only when both the momentum of the expansion coefficients and the space-time coordinates are restricted to the preferred direction. 
}
\end{table*}

\begin{figure*}
\hspace{-1cm}\includegraphics[scale=0.4]{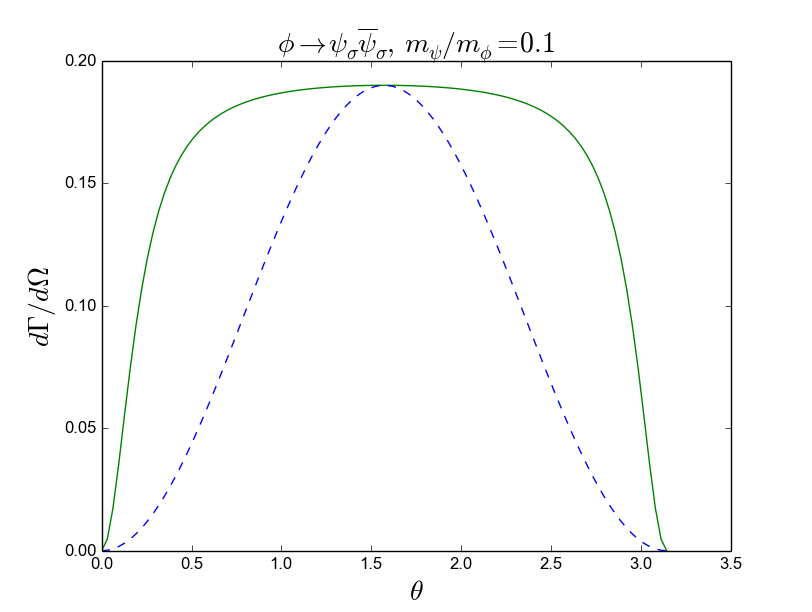}
\includegraphics[scale=0.4]{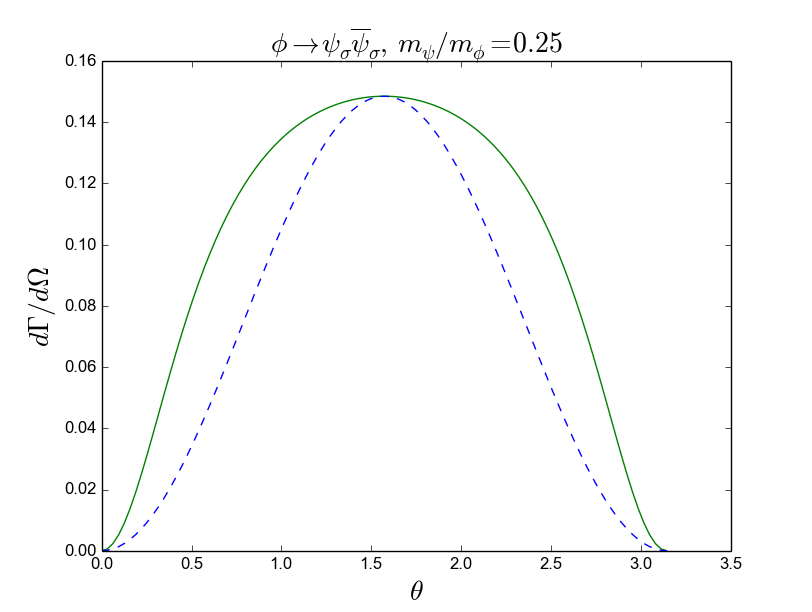}\\
\hspace{-1cm}\includegraphics[scale=0.4]{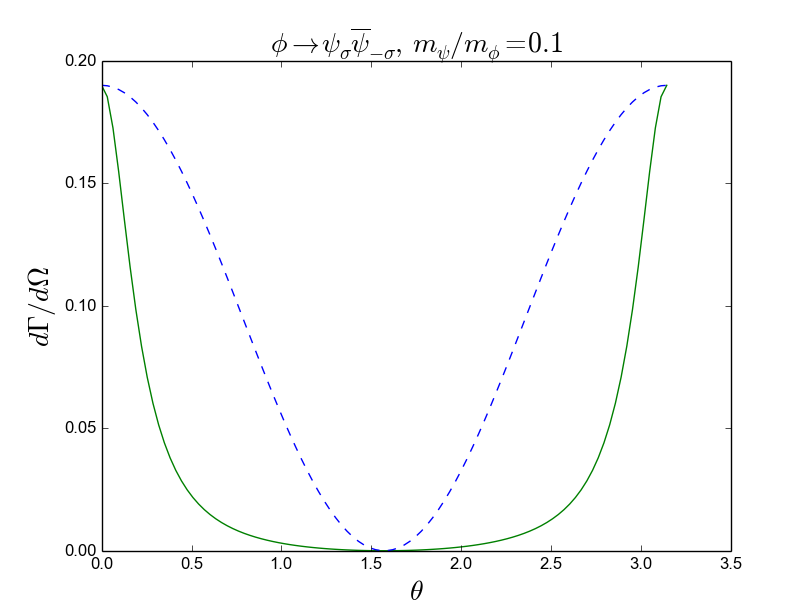}
\includegraphics[scale=0.4]{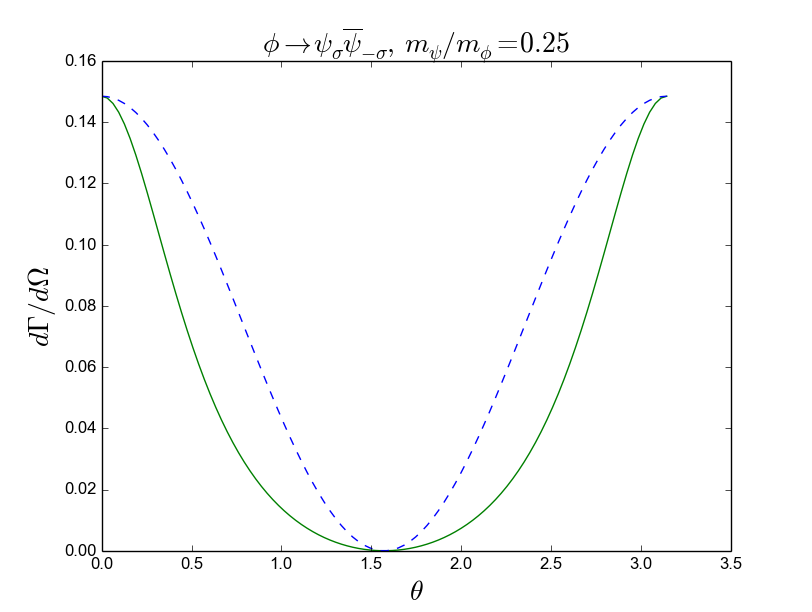}
\caption{The dotted and solid lines represent the differential polarised decay rate for SR and VSR respectively. The coupling is taken to be unity and the particle masses in the two sectors are identical. The bosonic mass is taken to be $m_{\phi}=125$ GeV.}\label{1}
\end{figure*}
\subsection{Interactions and the preferred direction} \label{sec:interaction}

A qualitative understanding of the effect of the preferred direction can be obtained by computing observables for physical processes. If we restrict ourselves to local interactions such as the standard gauge and Yukawa interactions, because the spin-sums between the Lorentz and VSR sectors are identical, all the unpolarised processes, apart from the coupling constants and the masses are the same. Therefore, the only difference are the polarised processes.

One of the simplest processes is the decay of a scalar boson into a pair of fermion anti-fermion pair described by the Yukawa interaction. In VSR, the differential decay rates for this process are given by
\begin{eqnarray}
\frac{d\Gamma_{\mbox{\tiny{VSR}}}}{d\Omega}(\phi\rightarrow\psi_{\sigma}\overline{\psi}_{\sigma})&=&
\frac{m_{\phi}g^{2}\sin^{2}\theta}{64\pi^{2}}
\left[1-\frac{4 m^{2}_{\psi}}{m^{2}_{\phi}}\right]^{3/2}
\left[\frac{(p^{0}_{\psi})^{2}-(p^{3}_{\psi})^{2}}{(p^{0}_{\psi})^{2}}\right]^{-1}\nonumber\\
&=&\frac{m_{\phi}g^{2}\sin^{2}\theta}{64\pi^{2}}
\left[1-\frac{4 m^{2}_{\psi}}{m^{2}_{\phi}}\right]^{3/2}
\left[1-\left(1-\frac{4m^{2}_{\psi}}{m^{2}_{\phi}}\right)\cos^{2}\theta\right]^{-1},
\end{eqnarray}
\begin{eqnarray}
\frac{d\Gamma_{\mbox{\tiny{VSR}}}}{d\Omega}(\phi\rightarrow\psi_{\sigma}\overline{\psi}_{-\sigma})&=&
\frac{m_{\phi}g^{2}\cos^{2}\theta}{64\pi^{2}}
\left[1-\frac{4 m^{2}_{\psi}}{m^{2}_{\phi}}\right]^{3/2}
\left[\frac{(p^{0}_{\psi})^{2}-(p^{3}_{\psi})^{2}}{m^{2}_{\psi}}\right]^{-1}\nonumber\\
&=&\frac{m_{\phi}g^{2}\cos^{2}\theta}{64\pi^{2}}
\left[1-\frac{4 m^{2}_{\psi}}{m^{2}_{\phi}}\right]^{3/2}
\left\{\frac{m^{2}_{\phi}}{4m^{2}_{\psi}}
\left[1-\left(1-\frac{4m^{2}_{\psi}}{m^{2}_{\phi}}\right)\cos^{2}\theta\right]\right\}^{-1}.
\end{eqnarray}
Comparing the above results to the same process with the VSR fermions replaced by the Dirac fermions, the difference are represented by the last terms of the respective equations which is the effects of the preferred direction. From fig.~\ref{1}, we see that as $m_{\psi}/m_{\phi}\rightarrow\frac{1}{2}$ the rates predicted by the two sectors coincide.

Since the kinematics of VSR fermions are identical to their Lorentz counterparts, the principles of local gauge invariance can be applied without any difficulties. The VSR fermions couple with gauge bosons through the conserved current $J^{\mu}=q\,\overline{\psi}\gamma^{\mu}\psi$. We find by explicit computation that the corresponding charge $Q=q\int d^{3}x \,J^{0}$ remain the same but the spatial component of the current $J^{i}(x)$ are different. Therefore, we should expect differences in the polarised processes. As for the VSR vector bosons, they inherit the same problem as their Lorentz-invariant counterparts, namely unitarity violation at high-energy. But in the VSR framework, since there are no physical massless vector bosons, the problem cannot be resolved using the Higgs mechanism.

\section{Conclusions}

Quantum field theory with VSR symmetry is an interesting subject that has received considerable attention since its conception. But to the best of our knowledge, an analysis based on the formalism of Wigner and Weinberg, has thus far not been carried out. In some cases, the starting point is a VSR-invariant but Lorentz-violating Lagrangian which modifies the kinematics of a Lorentz-invariant theory. Such an approach is unsatisfactory and unnecessary. Since we have knowledge of the underlying symmetry group and their representations, the field operators as shown in this paper, up to a global phase, are completely determined. To this end, we have shown that the VSR and Lorentz-invariant fields have the same locality structures and field equations. 

If we take the view that the VSR symmetry is a fundamental symmetry of nature, there is no reason stopping us from introducing additional VSR-invariant but Lorentz-violating interactions to the SM Lagrangians. This has been the main focus in the existing literature. In this paper, we have taken a different approach. The SM Lagrangian should remain unaltered and the VSR particles should be interpreted a new species of particles. Therefore, an important task is to incorporate them into an extended SM. Thus far, we have only studied the simplest interactions involving the VSR fermions. 

Confining ourselves to the VSR fermions, an important point to note is that since they have the same kinematics and locality structure as their Lorentz-invariant counterparts, the most natural interactions are the Yukawa and the standard gauge interactions. For these interactions, the effects of Lorentz violation are absent for unpolarised processes since the spin-sums between the two sectors are identical. The only differences are the polarised processes.


The hypothesis that in the low energy limit, the symmetry of nature is described by VSR  presents an intriguing paradigm due to the existence of a preferred direction. We have shown that one can obtain a qualitative understanding of its effect in a localised setting by studying polarised processes. The really interesting question is whether the preferred direction extends globally to the galactic or cosmic scale. If one proposes such an extension, then it seems to imply that the universe has a preferred direction due to some exotic non-trivial topology and there exist particles that are sensitive to the topology of the universe. To address this issue, one possibility is to extend general relativity by introducing appropriate VSR-invariant terms. 

In summary, the VSR framework provides a simple setting in which we can explore such possibilities. In our opinion, such a theory brings forward a new perspective on the notion of space-time. The advent of special relativity has taught us that space-time and simultaneity are relative and not absolute. The theories built upon the framework of VSR allow us to go one step further - the space-time symmetry according to different particles, may also be relative.

\section*{Acknowledgements}

I am grateful to D.~V.~Ahluwalia for various discussions in the past, suggesting that the existence of a preferred direction may imply a universe with non-trivial topology. This research is supported by the CNPq grant 313285/2013-6.


\bibliography{Bibliography}
 \bibliographystyle{unsrt}


\end{document}